\documentclass[preprint,showpacs,aps,preprintnumbers,amsmath,amssymb]{revtex4-1}

\usepackage{graphicx,epsfig}
\usepackage{dcolumn}
\usepackage{bm}
\newcommand{\be}{\begin{equation}}
\newcommand{\ee}{\end{equation}}
\newcommand{\bse}{\begin{subequations}}
\newcommand{\ese}{\end{subequations}}
\newcommand{\bary}{\begin{eqnarray}}
\newcommand{\eary}{\end{eqnarray}}
\newcommand{\bwt}{\begin{widetext}}
\newcommand{\ewt}{\end{widetext}}

\begin{document}


\title{EBL effect on the observation of multi-TeV flaring of 2009 from Markarian 501}
\author{Sarira Sahu }
\email{sarira@nucleares.unam.mx}
\author{Marco Vladimir Lemus Y\'a\~{n}ez} 
\email{vladimir@ciencias.unam.mx}
\author{Luis Salvador Miranda}
\email{luis.miranda@correo.nucleares.unam.mx}
\author{Alberto Rosales de Le\'{o}n}
\email{albertoros4@ciencias.unam.mx}
\author{Virendra Gupta$^a$}

\affiliation{
Instituto de Ciencias Nucleares, Universidad Nacional Aut\'onoma de M\'exico, 
Circuito Exterior, C.U., A. Postal 70-543, 04510 Mexico DF, Mexico\\
$^a$Departamento de  F\'isica Aplicada, 
Centro de Investigac\'ion y de Estudios Avanzandos del IPN\\
Unidad M\' erida, A. P. 73, Cordemex, M\'erida, Yucat\'an, 97310, M\'exico
}

\begin{abstract}
Markarian 501 is a high-peaked BL Lacertae object and has undergone
many major outburst since its discovery in 1996. As a part of the
multiwavelength campaign, in the year 2009 this blazar was observed for
4.5 months from March 9 to August 1 and during the period April 17 to May 5 it was observed
by both space and ground based observatories covering the entire
electromagnetic spectrum. A very strong high energy $\gamma$-ray
flare was observed on May 1 by Whipple telescope in the energy range 317 GeV to 5 TeV and the
flux was about 10 times higher than the baseline flux. We use the
photohadronic model complimented by the extragalactic background
radiation (EBL) correction to this very high state flare and have
shown that the EBL plays an important role in
attenuating the very high energy flux even though Markarian 501 is in
the local Universe. 

\end{abstract}
\maketitle

\section{Introduction}

Blazars are a
sub class of AGN and the dominant extra galactic population in gamma
rays\cite{Acciari:2010aa}. These objects show rapid variability in the entire
electromagnetic spectrum and have non thermal spectra which implies
that the observed photons originate within the highly relativistic
jets oriented very close to the observers line of sight\cite{Urry:1995mg}.
Due to the small viewing angle of the jet, it is possible to observe the
strong relativistic effects, such as the boosting of the emitted power
and a shortening of the characteristic time scales, as short as
minutes\cite{Abdo:2009wu,Aharonian:2007ig}. Thus these objects are important to study the energy
extraction mechanisms from the central super-massive black hole, 
physical properties of the astrophysical jets,
acceleration mechanisms of the charged particles in the jet and production of
ultra high energy cosmic rays, very high energy $\gamma$-rays and neutrinos.

The spectral energy distribution (SED) of these blazars
have a double peak structure in the $\nu-\nu F_{\nu}$ plane. 
The low  energy peak corresponds to
the synchrotron radiation from a population of relativistic electrons
in the jet and the high energy
peak believed to be due to the synchrotron self
Compton (SSC) scattering of the high energy electrons with their
self-produced synchrotron photon\cite{Dermer:1993cz,Sikora:1994zb}. Depending on the location of the first peak,
blazars are often sub classified into low energy peaked
blazars (LBLs) and high energy peaked blazars (HBLs)\cite{Padovani:1994sh}. In LBLs, the
first peak is in the near-infrared/optical energy range and the second peak is around GeV
energy range. For HBLs, the first peak is 
in the UV or X-rays range and the second peak is in the GeV-TeV energy range.
The above scenario is called leptonic model and  is very successful in explaining
the multi wavelength emission from blazars and FR I
galaxies\cite{Fossati:1998zn,Ghisellini:1998it,Abdo:2010fk,Roustazadeh:2011zz}.

Flaring seems to be the major activity of the blazars
which is unpredictable and switches between quiescent and
active states involving different time scales. 
While in some blazars a strong temporal correlation
between X-ray and multi-TeV $\gamma$-ray has been observed,
outburst in some other have no low energy counterparts (orphan
flaring)\cite{Krawczynski:2003fq,Blazejowski:2005ih} and explanation of such extreme activity
needs to be addressed through different mechanisms. It is also very
important to have simultaneous multiwavelength observation of the
flaring period to constraint different theoretical models of emission
in different energy regimes.

The TeV photons of the flare can interact with the background soft
photons in the jet to produce $e^+e^-$
pairs. However, production of the lepton pair within the jet depends on the 
size of the emitting region and the photon density in it. Also the  
required target soft photon threshold 
energy $\epsilon_{\gamma} \ge 2 m^2_e/E_{\gamma}$ is needed. 
It is observed that the jet medium is transparent to pair production where
the optical depth is very small\cite{Sahu:2013ixa,Abdop:2010}.
Also the TeV photons on their way to Earth can interact with  the extragalactic
background light (EBL) to produce the lepton pair\cite{Hauser:2001xs}.
However, TeV photons from the sources in the cosmologically local Universe (low redshift
sources) are believed to propagate unimpeded by the EBL, 
although the effect is found to be non negligible\cite{Abdop:2010}.

\section{Markarian 501}
Markarian 501 (Mrk 501)  (RA:$251.46^{\circ}$, DEC:$39.76^{\circ}$) is a HBL at a redshift of
z=0.034 (local Universe) is one of the brightest extragalactic sources in
X-ray/TeV sky\cite{Abdop:2010}. It is also the second extragalactic object (after
Markarian 421) identified as a very high energy (VHE) emitter by
Whipple telescope in 1996. Since its discovery, the multiwavelength
correlation of Mrk 501 have been studied intensively and during this
period it  has undergone many major outbursts on long time scales and
rapid flares on short times scales mostly in the X-rays and TeV
energies\cite{Pian:1998hh,Krawczynski:1999vz,Tavecchio:2001,Ghisellini:2002ex,Sambruna:2000ic,Gliozzi:2006qq,Villata:1999,Katarzynski:2001}. It has been observed that, during these outbursts, both the
peaks have shifted to higher energies and during the most extreme case
the synchrotron peak $\sim$ keV range has shifted above 200 keV\cite{Acciari:2010aa}. Due
to the low sensitivity of the previous generation instruments, Mrk 501
was primarily observed in VHE band during the outbursts. However,
later on it was observed in all the wave bands. In the year 2009, as a
part of large scale multiwavelength campaign covering a period of 4.5
months (from March 9 to August 1, 2009), Mrk 501 was observed\cite{Aliu:2016kzx}.
The scientific goal of this extended observation was to collect a
simultaneous, complete multifrequency data set to test the
current theoretical models of broadband blazar emission mechanism. Also
this will help to understand the the origin of high energy emission
from blazars and the physical mechanism responsible for the
acceleration of the charged particles in the relativistic jets. 
Between April 17 to May 5, Mrk 501 was observed by  both 
space and ground based observatories, covering the entire
electromagnetic spectrum including even the variation in optical
polarization\cite{Aliu:2016kzx}. A very strong VHE flare was detected first by Whipple
telescope on May 1st and 1.5 hours later with VERITAS. Both these
telescopes continued simultaneous observation of this VHE flare until
the end of the night. The detected flux
enhanced by a factor of $\sim10$ than the average baseline flux
($3.9\times 10^{-11}\, \text{ph}\,\text{cm}^{-2}\, \text{s}^{-1}$). A dramatic increase in
the flux by a factor $\sim 4$ in 25 minutes and a falling time of $\sim
50$ minutes was observed. The flux measured at lower energies before
and after the VHE flare did not show any significant 
variation. But, {\it Swift}-XRT (in X-ray) and UVOT (in optical) 
did observe moderate flux variability\cite{Aliu:2016kzx}. Also both Whipple and VERITAS
did observe statistically significant variation in VHE band.
Using the one-zone SSC model, the average
SED of this multiwavelength campaign of Mrk 501 is interpreted satisfactorily. 

Our aim here is to use the photohadronic model  of Sahu et al.\cite{Sahu:2013ixa,Sahu:2012wv,Sahu:2013cja,Sahu:2015tua,Sahu:2016bdu} and the EBL model of
Dominguez et al.\cite{Dominguez:2013lfa} to interpret the observed very strong VHE 
flare data of May 1. We found that this flare can be explained well
with this model.

\section{TeV flaring Model}
The photohadronic model of Sahu et al.\cite{Sahu:2013ixa,Sahu:2012wv,Sahu:2013cja}  rely on 
the standard interpretation of the leptonic model to explain both, low and high 
energy peaks, by synchrotron and SSC photons respectively as in the
case of any other AGNs and Blazars. 
Thereafter, it is proposed that the flaring 
occurs within a  compact and confined
volume of radius $R'_f$ inside the blob of radius $R'_b$ ($R'_f <
R'_b$)\cite{Sahu:2013ixa} (henceforth $^{\prime}$ implies the jet
comoving frame). 
Both the internal and the external jets are moving with the
same bulk Lorentz factor 
$\Gamma$ and the Doppler factor ${\cal  D}$ as the blob (for blazars
$\Gamma\simeq {\cal  D}$). In normal
situation within the jet, we consider the injected spectrum of the
Fermi accelerated charged particles having a power-law spectrum 
$dN/dE\propto
E^{-\alpha}$ with the power index $\alpha \ge 2$. But in the flaring
region the 
injected proton spectrum is a power-law spectrum supplemented with
an exponential decay factor and is given as
\be
\frac{dN_p}{dE_p}\propto  E_p^{-\alpha} e^{-E_p/E_{p,c}}.
\label{powerlawexp}
\ee
Here the high energy proton has the cut-off energy $E_{p,c}$.

 The high energy
protons will interact in the flaring region where the comoving photon
number density is $n'_{\gamma,f}$ to produce the $\Delta$-resonance.
Subsequently the $\Delta$-resonance decays to
charged and neutral pions and the further 
decay of neutral pions to TeV photons gives the multi-TeV SED. 
The $n'_{\gamma,f}$ is much higher than
the rest of the blob $n'_{\gamma}$ (non-flaring) i.e.
${n'_{\gamma, f}(\epsilon_\gamma)}\gg
{n'_{\gamma}(\epsilon_\gamma)}$. There is no direct way to estimate
the photon density in the inner jet region as it is hidden. For
simplicity we assume the scaling behavior of the photon densities in
the inner and the outer jet region as
\be
\frac{n'_{\gamma, f}(\epsilon_{\gamma_1})}
{n'_{\gamma, f}(\epsilon_{\gamma_2})} \simeq \frac{n'_\gamma(\epsilon_{\gamma_1})}
{n'_\gamma(\epsilon_{\gamma_2})},
\label{densityratio}
\ee 
which assumes that
the ratio of photon densities at two different
background energies $\epsilon_{\gamma_1} $  and $\epsilon_{\gamma_2} $
in flaring and non-flaring states remains almost the same. While the
photon density in the outer region can be calculated from the observed
flux, using Eq. (\ref{densityratio}) we can express the
$n'_{\gamma,f}$ in terms of $n'_{\gamma}$.

The $\pi^0$-decay TeV photon energy
$E_{\gamma}$ and the 
target SSC photon energy $\epsilon_{\gamma}$ in the observer frame are
related through,
\be
E_\gamma \epsilon_\gamma \simeq 0.032~\frac{ {\cal D}^2}{(1+z)^2} ~{\rm GeV}^2.
\label{Eegamma}
\ee
The observed TeV $\gamma$-ray energy and the proton energy
$E_p$ are related through
\be
E_p=\frac{10\Gamma}{\cal D} E_{\gamma}\simeq 10\,E_{\gamma}.
\label{Eproton}
\ee
The
optical depth of the $\Delta$-resonance process in the inner jet region
is given by
\be
\tau_{p\gamma}=n'_{\gamma, f} \sigma_{\Delta} R'_f,
\label{optdepth}
\ee
where the resonant cross section is $\sigma_{\Delta}\sim 5\times 10^{-28}\, cm^2$.
The efficiency of the  $p\gamma$ process depends on the
physical conditions of the interaction region, such as the size, 
the distance from the base of the jet, the photon density and their
distribution in the region of interest.

In the inner region we  compare the dynamical time scale $t'_{d}=R'_f$ 
with the $p\gamma$ interaction time scale
$t'_{p\gamma}=(n'_{\gamma,f}\sigma_{\Delta} K_{p\gamma})^{-1}$ to
constraint the seed photon density so that multi-TeV photons can be
produced. For a moderate efficiency of this process, we can assume
$t'_{p\gamma} > t'_{d}$ and this gives
$\tau_{p\gamma} < 2$, where the inelasticity parameter is assigned the
usual value of 
$K_{p\gamma}=0.5$. 
Also by assuming the Eddington luminosity is equally shared by the jet
and the counter jet, the luminosity within the inner region for a seed
photon energy $\epsilon'_{\gamma}$ will satisfy $(4\pi n'_{\gamma,f}
R'_f \epsilon'_{\gamma}) \ll L_{Edd}/2$.  This puts an upper limit on
the seed photon density as 
\be
n'_{\gamma,f}\ll \frac{L_{Edd}} {8\pi R'^2_f
\epsilon'_{\gamma}}.
\label{nedd}
\ee
From Eq.(\ref{nedd})  we can estimate the photon density in this
region. In terms of SSC photon energy and its luminosity, the photon number density
$n^{\prime}_{\gamma}$ is expressed as
\be
n^{\prime}_{\gamma}(\epsilon_{\gamma}) = \eta \frac{L_{\gamma,
    SSC} (1+z)}{{\cal D}^{2+\kappa} 4\pi {R^{\prime}}^2_b
  \epsilon_{\gamma}},
\label{nblob}
\ee
where $\eta$ is the efficiency of SSC process and $\kappa$ describes
whether the jet is continuous ($\kappa=0$) or discrete ($\kappa=1$). In this
work we take $\eta=1$ for 100\% efficiency.
The SSC photon luminosity  is expressed in terms of the observed flux
 ($\Phi_{SSC}(\epsilon_{\gamma}) =\epsilon^2_{\gamma} dN_{\gamma}/d\epsilon_{\gamma}$) and is
given by
\be
L_{\gamma,SSC}=\frac{4\pi d^2_L
  \Phi_{SSC}(\epsilon_{\gamma})}{(1+z)^2}.
\label{lssc}
\ee
Using the Eqs. (\ref{nblob}) and (\ref{lssc})  we can simplify the
ratio of photon densities given in Eq.(\ref{densityratio}) to
\be
\frac{n'_\gamma(\epsilon_{\gamma_1})}
{n'_\gamma(\epsilon_{\gamma_2})}=\frac{\Phi_{SSC}(\epsilon_{\gamma
    1})}{\Phi_{SSC}(\epsilon_{\gamma 2})}
 \frac{E_{\gamma_1}}{E_{\gamma_2}}.
\label{denratio}
\ee
The $\gamma$-ray flux from the $\pi^0$ decay is deduced to be 
\be
F_{\gamma}(E_{\gamma}) \equiv E^2_{\gamma} \frac{dN(E_\gamma)}{dE_\gamma} 
\propto  E^2_p \frac{dN(E_p)}{dE_p} n'_{\gamma,f} .
\ee
The exponential factor in the power spectrum in Eq. (\ref{powerlawexp})
is responsible for the decay of the VHE flux, and falls faster for
$E_{\gamma} > E_{c}$. Here $E_{c}$ is the $\gamma$-ray cut-off
energy corresponding to $E_{p,c}$.
The EBL effect also attenuates the VHE flux by a factor of
$e^{-\tau_{\gamma\gamma}}$, where $\tau_{\gamma\gamma}$ is the optical
depth which depends on the energy of the propagating VHE $\gamma$-ray and the
redshift $z$ of the source. So there is a competition between the
exponential cut-off and the EBL effect. It is well known that for
intermediate and large redshift objects, EBL plays a dominant role in
depleting the multi-TeV flux. However, for objects in the local
Universe (e.g. Mrk 421 and Mrk 501), it may not be important, although 
the multi-TeV flare data of Mrk 501 observed by MAGIC and VERITAS
telescopes were corrected for EBL effect. A 6 TeV photon was observed
during the 4.5 months campaign and the attenuation factor
$e^{-\tau_{\gamma\gamma}}$ for this photon is about $.4-0.5$ which is not
negligible\cite{Abdop:2010}. So here we would like to study the effect of EBL on the
strongest VHE flare of May 1 and compare with the exponential cut-off scenario.

Including the EBL effect, the relation between observed flux
$F_{\gamma}$ and the intrinsic flux $F_{int}$ is given as
\be
F_{\gamma}(E_{\gamma}) = F_{int}(E_{\gamma})
e^{-\tau_{\gamma\gamma}(E_{\gamma},z)}.
\label{fluxrelat}
\ee
Then the EBL corrected observed multi-TeV photon flux from $\pi^0$-decay
at two different observed photon energies $E_{\gamma 1}$ and
$E_{\gamma 2}$ can be expressed as
\be
\frac{F_\gamma(E_{\gamma_1})}{F_\gamma(E_{\gamma_2})} 
=
\frac{\Phi_{SSC}(\epsilon_{\gamma_1})}{\Phi_{SSC}(\epsilon_{\gamma_2})}
\left(\frac{E_{\gamma_1}}{E_{\gamma_2}}\right)^{-\alpha+3}
e^{-\tau_{\gamma\gamma}(E_{\gamma_1},z)+\tau_{\gamma\gamma}(E_{\gamma_2},z)},
\label{sscspectrum}
\ee
where we have used 
\be
\frac{E_{p_1}}{E_{p_2}}=\frac{E_{\gamma_1}}{E_{\gamma_2}}.
\ee
The $\Phi_{SSC}$ at different energies are calculated
using the leptonic model. Here the multi-TeV flux is proportional to
$E_{\gamma}^{-\alpha+3}$  and $\Phi_{SSC}(\epsilon_{\gamma})$. 
In the photohadronic process ($p\gamma$), the
multi-TeV photon flux is expressed as
\be
F(E_{\gamma})=A_{\gamma} \Phi_{SSC}(\epsilon_{\gamma} )\left (
  \frac{E_{\gamma}}{TeV}  \right )^{-\alpha+3}
e^{-\tau_{\gamma\gamma}(E_{\gamma},z)}.
\label{modifiedsed}
\ee
Both $\epsilon_{\gamma}$ and $E_{\gamma}$ satisfy the condition given in
Eq.(\ref{Eegamma}) and the dimensionless constant 
$A_{\gamma}$ is given by
\be
A_{\gamma}=\left( \frac{F(E_{\gamma_2})}{\Phi_{SSC}(\epsilon_{\gamma2}
    )}\right ) \left (
  \frac{TeV}{E_{\gamma_2}}  \right )^{-\alpha+3}e^{\tau_{\gamma\gamma}(E_{\gamma_2},z)}.
\label{agamma}
\ee
Comparing Eqs. (\ref{fluxrelat}) and (\ref{modifiedsed}), the intrinsic
flux $F_{int}$ is given as
\be
F_{int}(E_{\gamma})=A_{\gamma} \Phi_{SSC}(\epsilon_{\gamma} )\left (
  \frac{E_{\gamma}}{TeV}  \right )^{-\alpha+3}.
\label{fint}
\ee
Using Eq. (\ref{modifiedsed}), we can calculate the EBL corrected
multi-TeV flux where $A_{\gamma}$ can be fixed from observed flare
data. We can calculate the Fermi accelerated high energy proton flux $F_p$
from the TeV $\gamma$-ray flux through the relation\cite{Sahu:2012wv}
\be
F_p(E_p)=7.5\times \frac{F_{\gamma}(E_{\gamma})}{\tau_{p\gamma}(E_p)}.
\ee
The optical depth $\tau_{p\gamma}$ is given in Eq.(\ref{optdepth}).
For the observed highest energy $\gamma$-ray $E_{\gamma}$ corresponding
to a proton energy $E_p$, the proton flux
$F_p(E_p)$ will be always smaller than the Eddington flux
$F_{Edd}$. This condition puts a lower limit on the optical depth of
the process and is given by
\be
\tau_{p\gamma} (E_p) > 7.5\times
\frac{F_{\gamma}(E_{\gamma})}{F_{Edd}}.
\label{optdepthvhe}
\ee
From the comparison of different times scales and from
Eq.(\ref{optdepthvhe}) we will be able to constraint the seed photon
density in the inner jet region.

\begin{table}
\centering
\caption{These parameters (up to $B'$) are taken from the one-zone synchrotron
model of ref. \cite{Aliu:2016kzx} which are used to fit the SED of
Mrk 421. The last two parameters are obtained from the best
fit to the observed Whipple high state flare data  in our model.} 
\label{tab1}
\begin{tabular*}{\columnwidth}{@{\extracolsep{\fill}}llll@{}}
\hline
\multicolumn{1}{@{}l}{Parameter} &Description & Value\\
\hline
$M_{BH}$ & Black hole mass\cite{Barth:2002jr} & $(0.9-3.5)\times 10^9 M_{\odot}$\\
z & Redshift & 0.034\\
$\Gamma$ &Bulk Lorentz Factor & 12\\
${\cal D}$& Doppler Factor & 12\\
$R^{\prime}_b$ & Blob Radius & $1.2\times 10^{16}$cm\\ 
$B^{\prime}$ &Magnetic Field & $0.03$ G\\ 
\hline
$R'_f$ &Inner blob Radius& $5\times 10^{15}$cm\\
$\alpha$ &Spectral index& $2.4$\\
\hline
\end{tabular*}
\end{table}

\begin{figure}[t!]
\vspace*{-0.1cm}
{\centering
\resizebox*{0.7\textwidth}{0.4\textheight}
{\includegraphics{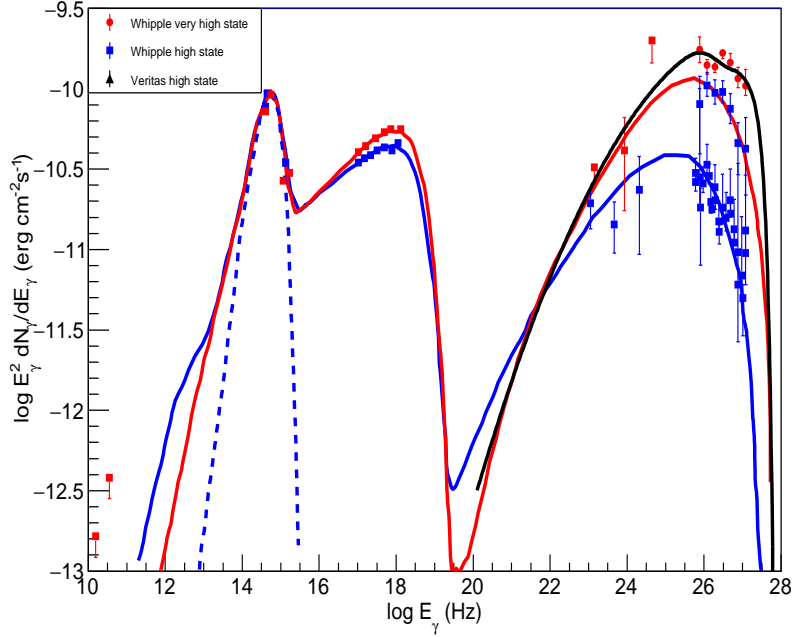}}
\par}
\vspace*{-0.30cm}
\caption{
The average SED of Mrk 501 is shown in all the energy bands which are
  taken from Ref. \cite{Aliu:2016kzx}. The SED of low state (MJD
  54936-54951; blue squares) and high state (MJD 54952-55; red
  circles) of the 3-week period are shown. The leptonic model fit to
  the low state (blue curve) and high state (red
  curve) are also shown. The blue dotted curve corresponds to the
  optical emission from the host galaxy. The black curve is the
  photohadronic fit to the Whipple very high state data (red circles). 
}
\label{mrk501sed}
\end{figure}

\begin{figure}[t!]
\vspace*{-0.1cm}
{\centering
\resizebox*{0.7\textwidth}{0.4\textheight}
{\includegraphics{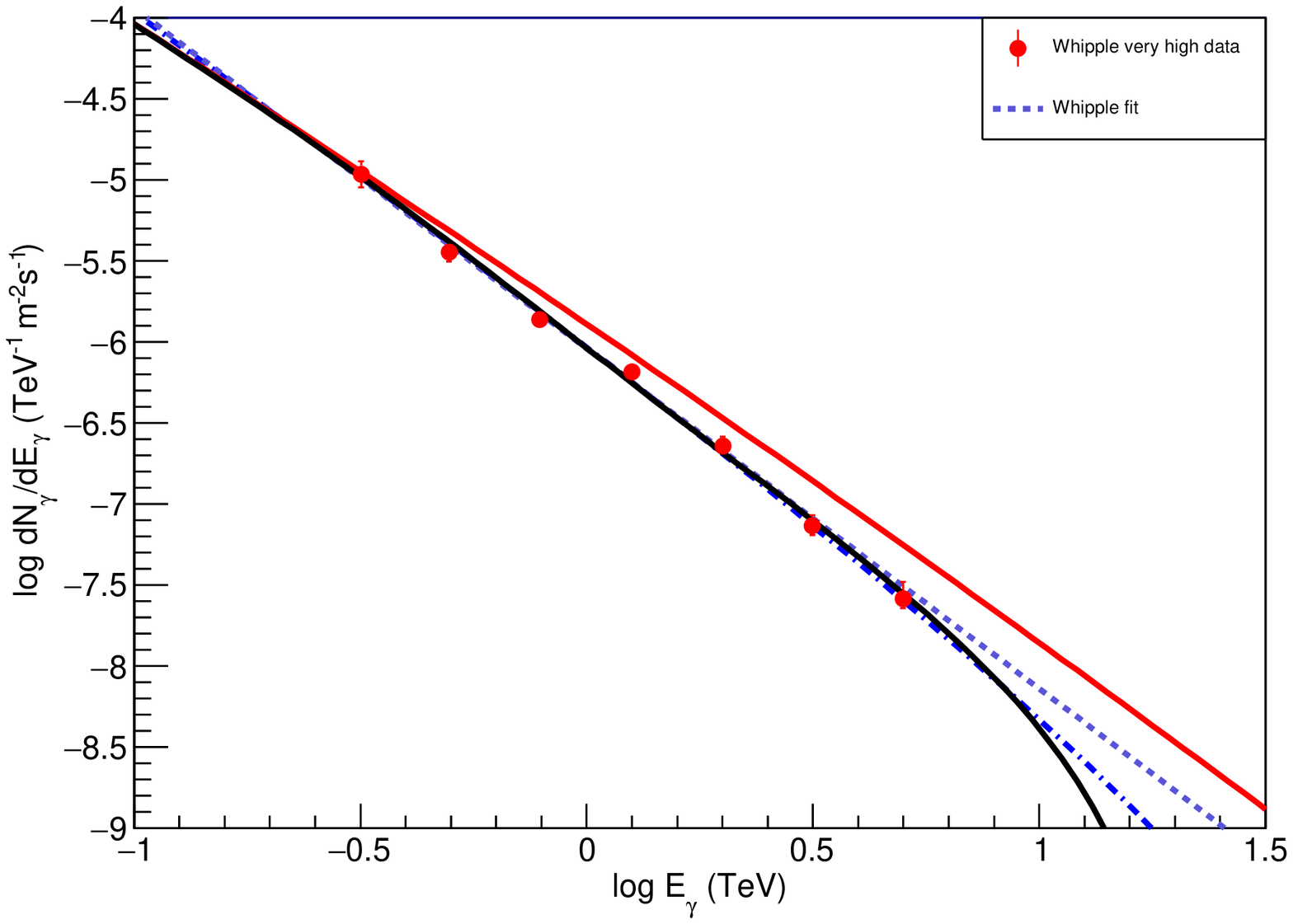}}
\par}
\vspace*{-0.30cm}
\caption{The black curve is the hadronic model fit to the
  Whipple very high state flare data (red filled circles) of Mrk 501
  and the red continuous curve is the intrinsic flux in the same
  model. For comparison we have also shown the Whipple fit to the data
  (dashed curve) and the exponential fit (dashed dotted curve).
}
\label{mrk501tevsed}
\end{figure}


\section{Results}

The average broadband SED of Mrk 501 is modeled using the standard
one-zone leptonic model\cite{Aliu:2016kzx}. The emission takes place from a spherical
blob of size $R^{\prime}_b$ which moves down the conical jet with a bulk Lorentz factor
$\Gamma$ and a Doppler factor ${\cal D}$. The emission region is filled with an isotropic and non-thermal
population of electrons and a randomly oriented magnetic field
$B^{\prime}$. To interpret the VHE flare of May 1, 2009, we use the
model parameters of the one-zone leptonic model which fits reasonably
well the average SED and the parameters are shown in Table \ref{tab1}.
 
The observed VHE flare of May 1, 2009 by Whipple telescope 
was in the range $\sim 317\,GeV \le E_{\gamma}\le\,5 \,TeV$. In the context of
photohadronic scenario, this range of $E_{\gamma}$ corresponds to the
Fermi accelerated proton energy  in the range $3.2\, TeV\le
E_p \le 50\, TeV$. So protons in this energy range will interact with
the background SSC photons in the energy range 
$13.6\, MeV (3.29\times 10^{21}\, Hz)\geq \epsilon_{\gamma} \geq\,\,
0.86 \, MeV  (2.1\times 10^{20}\, Hz)$
 to produce the
$\Delta$-resonance and subsequent decay of it will produce both
$\gamma$-rays and neutrinos through neutral and charged pion decay.
Also the above range of $\epsilon_{\gamma}$ lies in the beginning of
the SSC spectrum and in this range of energy
the sensitivity of the currently operating instruments are not good enough to detect Mrk
501. However, from the multiwavelength campaign the average
SED is fitted very well (Fig. \ref{mrk501sed}) and we use this low energy flux in the
 photohadronic model to calculate the observed flux.
Also to account for the
contribution of the EBL on the multi-TeV photons we consider the EBL
model by Dominguez et al. The EBL models of Dominguez et al.\cite{Dominguez:2013lfa} and
Franceschini et al.\cite{Franceschini:2008tp} are widely used to constraint the imprint of EBL
on the propagation of VHE $\gamma$-rays by Imaging Atmospheric
Cherenkov Telescopes (IACTs). The normalization constant $A_{\gamma}$ given in Eq. (\ref{agamma}) can be
calculated from the observed flare data. 

The multi-TeV flaring from blazars have an exponential fall which is
conventionally modeled as shown in Eq. (\ref{powerlawexp}). The cut-off energy $E_c$ is a free
parameter and depends on some  unknown mechanism.  
On the other hand, the
diffuse background radiation also  attenuate the high energy
$\gamma$-rays as a consequence of the lepton pair production.
In the local Universe EBL effect is assumed
to be very small. So in most of the flux calculation from the sources
in the local Universe, the EBL correction is neglected.   
However, here instead of the additional exponential cut-off, we take
into account the effect of EBL to deplete the intrinsic VHE flux.
A very good fit to the
Whipple very high state data of May 1 is obtained for $\alpha=2.4$ and
$A_{\gamma}=89$ where the EBL corrected flux is considered. We observed
that the EBL correction to the VHE
$\gamma$-ray is small but not insignificant (black curve in
Fig. \ref{mrk501tevsed}) 
and above 10 TeV it has a faster fall. We have also shown the
intrinsic flux (red curve in Fig. \ref{mrk501tevsed}) 
to demonstrate the difference. For comparison we have fitted the
data with an exponential cut-off function (dashed dotted curve) and the best fit is obtained for
$\alpha=2.6$, $E_c=30$ TeV  and $A_{\gamma}=66$. Also we have shown the 
Whipple fit (dashed curve) for comparison, where it is fitted by the
function $dN_{\gamma}/dE_{\gamma}=9.1\times 10^{-7} (E_{\gamma}/1 TeV)^{-2.1}\,
\text {ph}\, \text {m}^{-2}\, \text {s}^{-1}\, \text {TeV}^{-1}$. It is observed that
the very high state data of Whipple fits very well with above three
scenarios and all are same. However, above 5 TeV, both the EBL
corrected fit and the exponential fit differ from the Whipple
fit. Again the EBL fit and the exponential fit differ above 10 TeV and
the former one falls faster than the latter as can be seen from  Fig. \ref{mrk501tevsed}.
Even though all these fit very
well with the Whipple very high state data, we observe deviation in the VHE limit. 
So observation of the VHE flux above 10 TeV will be a good test to
constraint the EBL effect on the propagation of VHE $\gamma$-rays. In
Fig. \ref{mrk501sed}, we also plotted the Whipple very high state
data and our model fit (black curve) along with the
complete SED.

The high energy protons will be accompanied by high energy electrons and these
electrons will emit synchrotron photons in the energy range $\sim
10^{19}\, Hz$ to $\sim 10^{23}\, Hz$ when encountering the magnetic
field of the jet. This energy range photons lie in between the high
energy end of the synchrotron spectrum and the low energy tail of 
the SSC spectrum, thus may not be observed due to their low flux in
this region. These high energy electrons will also emit 
SSC photons and their energy is given by $E_{IC}\sim \gamma^2_e
\epsilon_{syn}$.

As discussed before, in the flaring state, in  general, the
flux of the two opposing jets can be as high as $F_{Edd}/2$. However,
the highest energy protons with $E_p=50$ TeV must have a flux $F_p <
F_{Edd}/2 \simeq 0.8 \times 10^{-7} \, \text{erg}\, \text{cm}^{-2}\, \text{s}^{-1}$. This constraint translates into
$\tau_{p\gamma} > 0.04$ which corresponds to $n'_{\gamma, f} > 1.5\times 10^{10}\,
cm^{-3}$ in the inner jet. However, the hidden jet lies between $R_s$
(Schwarzschild radius) and $R'_b$. As one representative
value we take $R'_f\simeq 5\times 10^{15}$. 
From Eq.(\ref{nedd})
the seed photon density for $\epsilon_{\gamma}=0.86$ MeV satisfies the inequality
$ n'_{\gamma,f} < 5.1\times 10^{10} cm^{-3}$ which translates to the 
optical depth to be constrained as $\tau_{p\gamma} < 0.13$. 
So the optical depth lie in the range $0.04 < \tau_{p\gamma} <
0.13$ and this corresponds to the range of photon density in the
inner jet region as $1.5\times 10^{10}\,
cm^{-3} < n'_{\gamma,f} < 5.1\times 10^{10} cm^{-3}$, which shows that the photon
density in this region is high. 
Due to the adiabatic expansion of the inner blob,
the photon density will be reduced to $n'_{\gamma}$ and also the
optical depth $\tau_{p\gamma} \ll\, 1$. The energy
will dissipate once these photons cross into the
bigger outer cone. This will drastically reduce the $\Delta$-resonance
production efficiency from the $p\gamma$ process.

\section{Conclusions}

The VHE flare of May 1, 2009 observed by Whipple telescopes can be
explained very well through photohadronic model supplemented with the
EBL correction. Previously, the decay of the VHE flare can
be explained through the exponential fall of the flux 
which introduces an additional free parameter, the cut-off
energy. However, here, the EBL corrected VHE flux automatically falls
exponentially without any additional free parameter and fits very well
with the Whipple very high state data. For comparison we
have also shown the Whipple fit as well as the exponential fit. All
these three curves fit very well with the VHE flare data. However, we have
shown that their behaviors differ in the high energy limit.
Observation of flare events above 10 TeV will be able to constraint 
different models and also shed more light on the EBL contribution to
the propagation of VHE $\gamma$-rays in the local Universe.


We are thankful to Lucy Fortson for many useful discussions. 
The  work of S.S. is partially supported by DGAPA-UNAM (M\'exico) Project
No. IN110815. V. Gupta is thankful to COCACYT (M\'exico) for partial support.

\end{document}